\documentclass[aps,amsmath,amssymb,reprint,superscriptaddress,preprintnumbers,showpacs,intlimits,longbibliography]{revtex4-1}
\pdfoutput=1
\bibpunct{[}{]}{,}{n}{}{}

\usepackage[dvipsnames]{xcolor}
\usepackage{latexsym,enumerate}
\usepackage{bm,mathrsfs}
\usepackage[cal=boondoxo]{mathalfa}

\usepackage[breaklinks=true,unicode=true,urlcolor = blue,colorlinks = true,citecolor = blue,linkcolor = blue]{hyperref}
\usepackage{mathtools}
\usepackage{cleveref}
\usepackage[labelformat=simple]{subcaption}

\usepackage{graphicx}
\usepackage{xifthen}
\usepackage{bigints}

\usepackage[toc,page]{appendix}
\graphicspath{{./}}
\renewcommand{\vec}[1]{\bm{#1}}

\DeclareMathOperator{\sech}{sech}

\begin{document}
	
\title{Engineered Chiral Skyrmion and Skyrmionium States by the Gradient of Curvature}
	
\author{Oleksandr V. Pylypovskyi}
\email{engraver@knu.ua}
\affiliation{Taras Shevchenko National University of Kyiv, 01601 Kyiv, Ukraine}
	
\author{Denys Makarov}
\email{d.makarov@hzdr.de}
\affiliation{Helmholtz-Zentrum Dresden-Rossendorf e.\,V., Institute of Ion Beam Physics and Materials Research, 01328 Dresden, Germany}

\author{Volodymyr~P.~Kravchuk}
\email{vkravchuk@bitp.kiev.ua}
\affiliation{Bogolyubov Institute for Theoretical Physics of National Academy of Sciences of Ukraine, 03680 Kyiv, Ukraine}
\affiliation{Leibniz-Institut f{\"u}r Festk{\"o}rper- und Werkstoffforschung, IFW Dresden, D-01171 Dresden, Germany}

\author{Yuri~Gaididei}
\email{ybg@bitp.kiev.ua}
\affiliation{Bogolyubov Institute for Theoretical Physics of National Academy of Sciences of Ukraine, 03680 Kyiv, Ukraine}

\author{Avadh Saxena}
\email{avadh@lanl.gov}
\affiliation{Theoretical Division, Los Alamos National Laboratory, Los Alamos, New Mexico 87545, USA}

\author{Denis D. Sheka}
\email{sheka@knu.ua}
\affiliation{Taras Shevchenko National University of Kyiv, 01601 Kyiv, Ukraine}

\date{July 5, 2018}

%
%

\begin{abstract}
Curvilinear nanomagnets can support magnetic skyrmions stabilized at a local curvature without any intrinsic chiral interactions. Here, we propose a new mechanism to stabilize chiral N\'{e}el skyrmion states relying on the \textit{gradient} of curvature. We illustrate our approach with an example of a magnetic thin film with perpendicular magnetic anisotropy shaped as a circular indentation. We show that in addition to the topologically trivial ground state, there are two skyrmion states with winding numbers $\pm 1$ and a skyrmionium state with a winding number $0$. These chiral states are formed due to the pinning of a chiral magnetic domain wall at a bend of the nanoindentation due to spatial inhomogeneity of the curvature-induced Dzyaloshinskii--Moriya interaction. The latter emerges due to the gradient of the local curvature at a bend. While the chirality of the skyrmion is determined by the sign of the local curvature, its radius can be varied in a broad range by engineering the position of the bend with respect to the center of the nanoindentation. We propose a general method, which enables one to reduce a magnetic problem for any surface of revolution to the common planar problem by means of proper modification of constants of anisotropy and Dzyaloshinskii--Moriya interaction.
\end{abstract}


\maketitle


\section{Introduction}
\label{sec:intro}

Chiral magnetic textures as domain walls, skyrmions and skyrmion bubbles are considered as promising building blocks for prospective memory and logic devices relying on spintronics and spino rbitronics concepts \cite{Nagaosa13,{Finocchio16},Wiesendanger16,{Fert17}}. There is an intensive work on the controlled creation of these topologically non-trivial objects~\cite{Seidel16,Fert17,Sapozhnikov16,Sapozhnikov15,Sapozhnikov15a,Gallagher17} but also on the manipulation of their static and dynamic properties~\cite{Seidel16,Fert17,Psaroudaki18,Mueller17}. Primarily, activities are dedicated to flat magnetic thin films with perpendicular magnetic anisotropy. Recently, it was shown that local curvature can lead to the emergent exchange-driven Dzyaloshinskii--Moriya interaction (DMI)~\cite{Gaididei14,Sheka15} enabling the route to realize skyrmions~\cite{Kravchuk16a} and field-free skyrmion lattices even as a ground state~\cite{Kravchuk18a}. 

Here, we demonstrate that, even in the absence of an intrinsic DMI, the gradient of the local curvature is an efficient mean to \textit{stabilize} chiral localized magnetic objects allowing to \textit{manipulate} their size at will. We pinpoint the physical mechanism of the effect to be the pinning of a chiral magnetic domain wall on an inhomogeneity of the geometry-driven DMI localized at the bend of a ferromagnetic nanomembrane. Engineering the geometry of a circular nanoindentation~[Fig.~\ref{fig:intro}(a)] to have a defined curvature and distance between bends allows to form chiral objects with winding numbers $Q$ of $\pm 1$ and $0$. Considering their topological properties, we refer to the objects with $Q=\pm 1$ as skyrmion states and $0$ as skyrmionium state~\cite{Finazzi13,Komineas15c}. The diameter of a skyrmion is determined by the diameter of the circular base of the nanoindentation. The developed theoretical formalism allows to transfer the conclusions to flat systems with spatially inhomogeneous DMI. In this respect, we propose a new mechanism of pinning of magnetic domain walls on gradients of DMI in a film.

\section{Results}
\label{sec:results}

\begin{figure*}
\includegraphics[width=\linewidth]{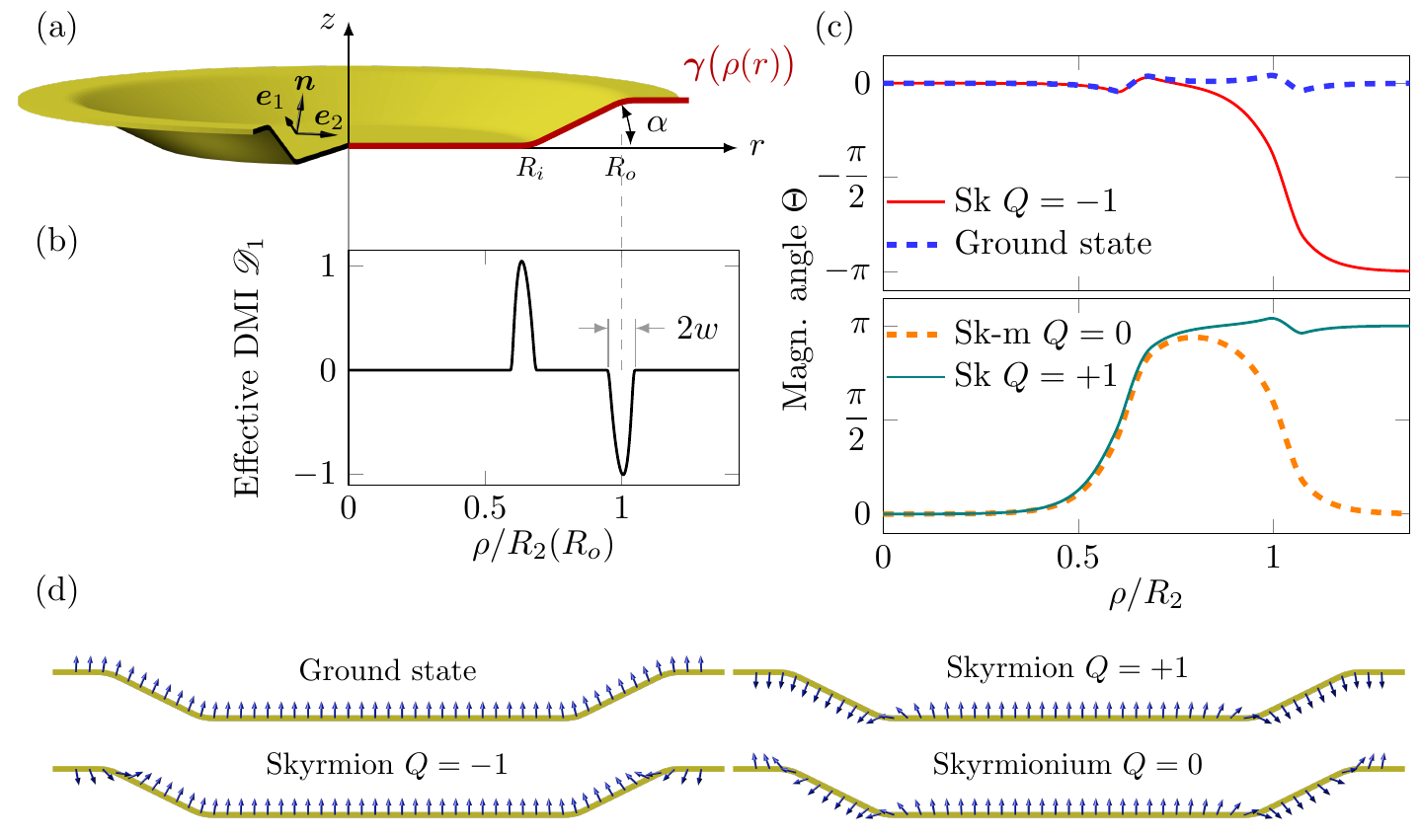}
\caption{\textbf{Sample geometry and magnetization patterns.} (a) Circular nanoindentation with inner radius $R_i = 10$, outer radius $R_o = 15$ and bending angle $\alpha = 26^\circ$ (curvature amplitude $\varkappa_0 = 0.5$ and curvature half-width $w = 0.75$, see Appendix~\ref{app:sk-stab} for details). A local basis is given by radial, polar and normal unit vectors $\{\vec{e}_1, \vec{e}_2, \vec{n}\}$, the cone angle equals $\pi-\alpha$. (b) The effective DMI coefficient $\mathscr{D}_1$ plotted as function of $\rho$, see Eq. \eqref{eq:D1-D2}. (c) Equilibrium states: topologically trivial ground state and three types of skyrmions of different chiralities (winding number $Q = \pm 1$) and a topologically trivial skyrmionium state ($Q = 0$). (d) Magnetization distributions for states shown~in~(c).}
\label{fig:intro}
\end{figure*}

We consider an infinitely thin 3D curved magnetic nanomembrane of thickness $h$ with uniaxial, locally perpendicular anisotropy in the absence of an intrinsic DMI. The nanomembrane is curved in a way to form a circular nanoindentation, Fig.~\ref{fig:intro}(a). 

A magnetic texture is controlled by three interactions: exchange, requesting uniform magnetization in the laboratory reference frame; anisotropy, tracking the nanomembrane curvature, and local magnetostatics. The competition between the first two results in the characteristic magnetic length $\ell = \sqrt{A/K}$ with the exchange stiffness $A$ and effective anisotropy coefficient $K = K_u - 2\pi M_\textsc{s}^2$. The latter incorporates any intrinsic anisotropy $K_u$ along the normal $\vec{n}$ to the surface and local magnetostatics in the thin film limit $h \lesssim \ell$~\cite{Gioia97,Carbou01,Kohn05a,Fratta16b}, see also Eq.~\eqref{eq:energy-total}. 

The nanoindentation geometry is defined by the revolution of a curve $\vec{\gamma}(r) = r \hat{\vec{x}} + z(r)\hat{\vec{z}}$ around $\hat{\vec{z}}$ axis resulting in a surface $\vec{\varsigma}$, see Fig.~\ref{fig:intro}(a) and Appendix~\ref{app:surf-rev} for details. It represents a conic frustum indentation with an inner radius $r = R_i$ and outer radius $r = R_o$. Here and below, all lengths are measured in units of $\ell$. We will characterize $\vec{\varsigma}$ through its two principal curvatures $\varkappa_1(r)$ and $\varkappa_2(r)$, where $\varkappa_1$ is the normal curvature of the generatrix $\vec{\gamma}$.

Surfaces of revolution support radially symmetric magnetization textures $\vec{m}(r) = \vec{M}/M_\textsc{s} = \sin\Theta \vec{e}_1 + \cos\Theta \vec{n}$ with $M_\textsc{s}$ being the saturation magnetization, see Appendix~\ref{app:curv-energy} for details. Here, the local orthonormal reference frame $\{\vec{e}_1, \vec{e}_2, \vec{n} \}$ is used with $\vec{e}_1$ being the unit vector along the generatrix, $\vec{e}_2 = \vec{n} \times \vec{e}_1$ and $\Theta = \Theta(r) \in \mathbb{R}$. 

To compare topologically nontrivial magnetization textures in flat and curvilinear samples, it is instructive to project $\vec{\varsigma}$ to a plane in such a way to reconstruct a planar skyrmion equation, see, e.g. Eq.~(13) from the Ref.~\cite{Leonov16}. We introduce a surface polar coordinate
\begin{equation}\label{eq:rho-via-r}
\rho(r) = r \exp \bigintsss_r^\infty \left[1-\sqrt{1+\left(\frac{\mathrm dz}{\mathrm d\zeta}\right)^2}\right]\dfrac{\mathrm d\zeta}{\zeta}.
\end{equation}
We note that far from the center of the nanoindentation $\rho(r) = r$, see Fig.~\ref{fig:rho-plot}(a). Nanomembrane bends in new coordinates are located at $\rho(R_i) = R_1$ and $\rho(R_o) = R_2$. Taking into account the locally perpendicular anisotropy, the radially symmetric magnetization texture is given by a forced skyrmion equation
\begin{subequations} \label{eq:forced-skyrmion}
\begin{align} \label{eq:static-equation}
&\!\! \Theta'' + \frac{\Theta'}{\rho} - \frac{\sin 2\Theta}{2\rho^2}- \frac{\mathscr{K}}{2}\sin 2\Theta - \mathscr{D}_2 \frac{\sin^2\Theta}{\rho} \!=\! f(\rho),\shortintertext{and}
\label{eq:force}
&f(\rho) = \frac{\mathscr{D}_1 - \mathscr{D}_2}{2\rho} + \frac{1}{2} \mathscr{D}_1' = \frac{r(\rho)}{\rho} \left(\varkappa_1 + \varkappa_2\right)'
\end{align}
\end{subequations}
with prime denoting the derivative with respect to $\rho$, see Appendix~\ref{app:curv-energy} for details.
The boundary condition in the origin $\Theta(0) = 0$. The value of $\Theta(\infty) = Q\pi$ gives a winding number of the magnetization through directions $\vec{n}$ and $-\vec{n}$ (skyrmion chirality) $Q = \pm 1$. The case of $Q = 0$ can be either skyrmionium state ($Q=0$)~\cite{Finazzi13,Komineas15c} or topologically trivial state. Four solutions of~\eqref{eq:forced-skyrmion} with different $Q$ are shown in Fig.~\ref{fig:intro}(c,d) and will be discussed below.

The effective anisotropy $\mathscr K$ and DMI $\mathscr{D}_2$ in Eq.~\eqref{eq:static-equation} are functions of the coordinate $\rho$. In the case $\mathscr{K}=\text{const}$, $\mathscr{D}_2=\text{const}$, and $f(\rho)=0$, Eq.~\eqref{eq:forced-skyrmion} is reduced to a typical skyrmion-like equation~\cite{Bogdanov94,Rohart13,Leonov16,Seidel16}. The key difference of our work [Eq.~\eqref{eq:forced-skyrmion}] to the standard skyrmion equation is the presence of the  spatial dependence of the anisotropy and DMI parameters and an external driving with a spatial-dependent external force $f(\rho)$ resulting in the absence of the strictly normal magnetization pattern. Both DMI coefficients are proportional to the corresponding principal curvatures of the nanomembrane~\cite{Gaididei14}, $\mathscr{D}_i \propto \varkappa_i$ with $i = 1,2$. 
Thus, $\mathscr D_1$ is nonzero only in the bend regions of the generatrix $\vec{\gamma}$ and $\mathscr{D}_2$ is nonzero only in an inclined part of the indentation. While $\varkappa_1$ can be of arbitrary value and is given by the generatrix bend parameters only, $\varkappa_2 \propto 1/r$. The effective anisotropy $\mathscr{K}$ is determined by $\varkappa_2$ and the ratio $r/\rho$. An example of the spatial dependency of the $\mathscr{D}_1$ is shown in Fig.~\ref{fig:intro}(b).

Based on Eq.~\eqref{eq:forced-skyrmion}, we obtain two distinct ways to control the type of the magnetic texture: (i)~Geometry-induced DMI in the left-hand side of Eq.~\eqref{eq:static-equation} resulting in the stabilization of small-radius skyrmions~\cite{Kravchuk16a}; (ii)~Effective external driving force originating from the inhomogeneity of a local curvature~\eqref{eq:force}. In the following, we show that the latter one allows for the formation of chiral skyrmions of \emph{tunable radii}. 

For a nanoindentation geometry, Fig.~\ref{fig:intro}(a), the ground state of the system is a quasi-normal magnetization distribution indicated with a blue solid line in Fig.~\ref{fig:intro}(c). 
The deviation of the local magnetization from the strictly normal direction is an exchange-driven effect, forcing the magnetization distribution to be homogeneous in the laboratory reference frame. 
For a slow varying curvature $\varkappa_1$ (i.e. $|\varkappa_1'| \ll 1$) and large indentation radius $R_{1,2}\gg 1$, one can omit the terms inversely proportional to powers of $R_{1,2}$ in Eq.~\eqref{eq:static-equation} and obtains $\Theta(\rho) \approx - \varkappa_1' r/\rho$.
Other possible solutions represent a N\'{e}el  skyrmion.
While small-radius skyrmions can appear for an arbitrary geometry-induced DMI with a radius governed by the DMI coefficient~\cite{Kravchuk16a,Kravchuk18a}, in the present case we obtain three magnetization textures with different winding numbers $Q$, Fig.~\ref{fig:intro}(c). Inner and outer bends with positive and negative signs of $\mathscr{D}_1$, respectively, support skyrmions with $Q = \pm 1$ and skyrmionium state with $Q = 0$.

\begin{figure}
	\includegraphics[width=\linewidth]{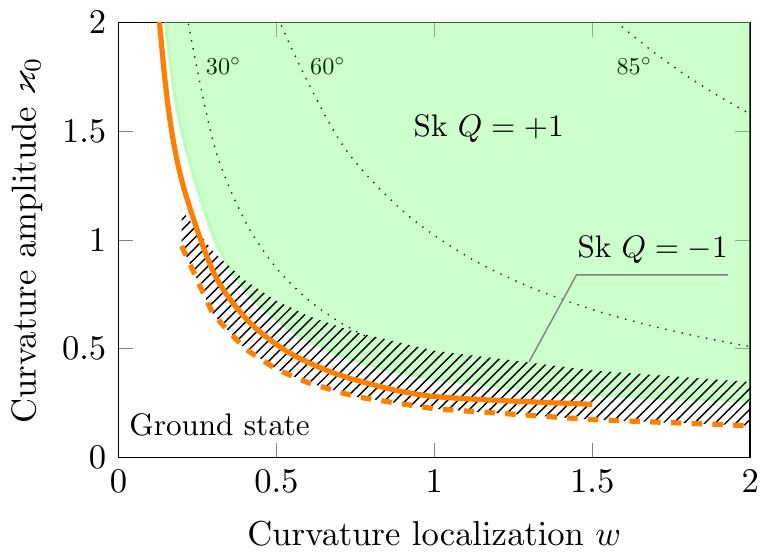}
	\caption{\textbf{Stability regions of skyrmions in coordinates of the curvature amplitude $\varkappa_0$ and curvature spatial localization $w$.}  Skyrmion with $C = +1$ pinned at the inner bend of the nanoindentation with $R_i = 10$ and $R_o = 50$ is stable in the green-shaded region. Orange line shows the corresponding asymptotics for a sharp bend~\eqref{eq:dw-stab} and three dotted curves are isolines for $\alpha$ of $30^\circ$, $60^\circ$ and $85^\circ$ [bend inclination angle, Fig.~\ref{fig:intro}(a)]. Skyrmion with $C = -1$ pinned at the outer bend of the nanomembrane with $R_i = 10$ and $R_o = 15$ is stable in the dashed region. Asymptotics~\eqref{eq:dw-stabo} is shown by dashed line. 
	}
	\label{fig:stability}
\end{figure}

Analytically, we consider a model with a sharp bell-shaped bend of half width $w \ll 1$ and curvature amplitude $\varkappa_0$, see Appendix~\ref{app:sk-stab} for details. Then, the curvature can be represented as 
\begin{equation}\label{eq:delta-curvature}
\varkappa_1(\rho) = \alpha \left[ \delta(\rho-R_1) - \delta(\rho-R_2) \right]
\end{equation}
with $\delta(\bullet)$ being the Dirac $\delta$-function and $\alpha > 0$ being the bending angle. Both radii of the indentation are assumed to be large, $R_2 > R_1 \gg1$. These assumptions lead to simplifications with $\mathscr{D}_2 = 0$ and $\mathscr{K} = 1$. 

Using an ansatz of a circular domain wall
\begin{equation}\label{eq:circular-dw}
\Theta_\text{sk} = 2\arctan \left[ p\exp (\rho-R_\text{sk}) \right] + (p-1)\pi/2
\end{equation} 
with $p = \pm 1$, the total energy reads
\begin{equation}\label{eq:skyrmion-en}
\mathcal{E} = 4R_\text{sk} - 2p \alpha \left[ R_1 \sech(R_\text{sk} - R_1) - R_2 \sech(R_\text{sk} - R_2) \right].
\end{equation}
Here, the first and second terms represent the energy of a circular domain wall of radius $R_\text{sk}$ and the contribution of the effective DMI, respectively, c.f. with Eq.~(5)  \cite{Kravchuk18}. If the domain wall is localized near $R_1$ and $R_2$, the stability conditions read
\begin{subequations} \label{eq:dw-stab-stabo}
\begin{align} \label{eq:dw-stab}
p\alpha > \dfrac{4}{R_1} & \quad\mbox{for the inner bend}, \\ 
\label{eq:dw-stabo} %
-p\alpha > \dfrac{4}{R_2} & \quad\mbox{for the outer bend},
\end{align}
\end{subequations}
see Appendix~\ref{app:sk-stab} for details. The coefficient $p$ is equal to $\pm 1$ for outward and inward magnetization rotation, respectively. Due to the different signs of the effective DMI in the inner and outer bends, skyrmions of different chiralities can be pinned. This is also the reason for the stabilization of a skyrmionium state with zero total winding, see orange dashed line in Fig.~\ref{fig:intro}(c). A wider skyrmion can be pinned at the first bend with a smaller bending angle $\alpha$. 

To model the curvature of a finite spatial localization, we choose the first principal curvature as a sum of two bell-shaped functions with the maximal value $\varkappa_0$ strongly localized in rings of radii $(R_i-w,R_i+w)$ and $(R_o-w,R_o+w)$. The curvature is zero outside these rings, see Appendix~\ref{app:sk-stab} for details. 

Fig.~\ref{fig:intro}(c) shows four solutions of Eq.~\eqref{eq:static-equation} for a concave nanoindentation [Fig.~\ref{fig:intro}(a)]: ground state (blue dashed line), two skyrmions of different signs of $Q$ (solid red and green lines), and skyrmionum state (orange dashed line). A magnetic domain wall is pinned near the maximum of the curvature and slightly shifted to the bottom flat side of the sample. The impact of a finite curvature is shown in Fig.~\ref{fig:stability}(a).

There are separate stability regions for skyrmions of different winding numbers $Q$. Skyrmions with $Q = +1$  pinned at the inner bend of the indentation are stable in a wide range of curvature amplitude $\varkappa_0$ and half-width $w$, see green-shaded area in Fig.~\ref{fig:stability}. Skyrmions with $Q = -1$ are stable at a narrow dashed-indicated area. The upper boundary of the stability region is related with the small enough distance between the inner and outer bend. An increase of the curvature influences the domain wall shape. When a critical value of $\varkappa_0$ is reached, the domain wall slides down completing the magnetization reversal to the ground state. The analytically predicted lower boundaries of the stability regions for the pinning in the inner and outer bends~\eqref{eq:dw-stab-stabo} closely coincide with the numerically calculated ones in a wide range of parameters, see solid and dashed orange lines in Fig.~\ref{fig:stability}. The different stability regions for skyrmions with $Q = \pm 1$ are related with the effective DMI in the inner and outer bend of the nanoindentation: sign of $\mathscr{D}_1$ selects the clockwise or counter-clockwise direction of magnetization winding.

\begin{figure*}
	\includegraphics[width=\linewidth]{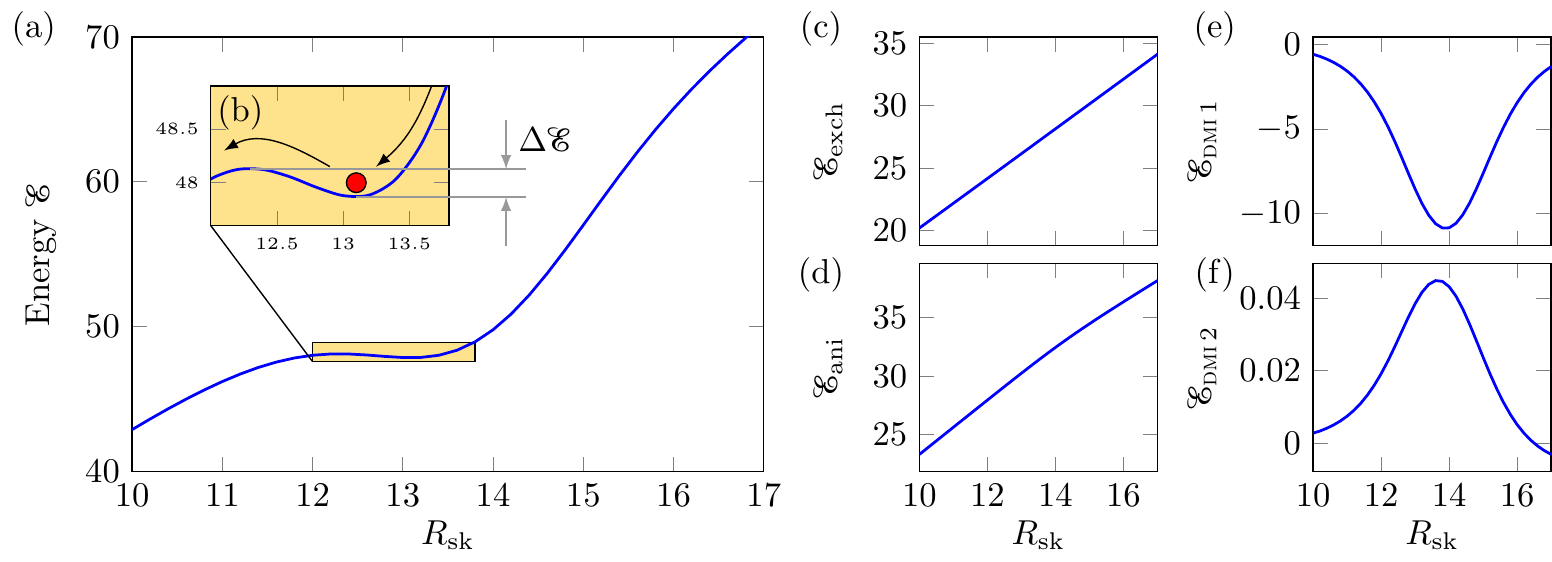}
	\caption{\textbf{Pinning of a skyrmion on the inner bend.} (a) Total energy of a skyrmion as a function of its radius $R_\text{sk}$ for $w = 1.5$ and $\varkappa_0 = 0.25$ with $R_i = 15$ (bending angle $\alpha = 26^\circ$). Energy profile has a minimum near $R_1(R_i) = 13.9$. (b) Magnified region of local minimum in $\mathcal{E}$.  Pinning is schematically shown with a red point moving from the right top corner and stopped in the local energy minimum. (c)--(f) Energy profiles of exchange, effective anisotropy and two effective DMI, see energy density~\eqref{eq:energy}. Only the energy  related with $\mathscr{D}_1$ in~(d) allows pinning, while $\mathcal{E}_{\textsc{dmi}\,2}$ is two orders smaller in magnitude for any finite $\mathscr{D}_2 \neq 0$.
	}
	\label{fig:pinning}
\end{figure*}

Using the ansatz~\eqref{eq:circular-dw} with $p = 1$, we estimate the energy profile~\eqref{eq:energy} and energy gap allowing a skyrmion to be pinned at a bend, see Fig.~\ref{fig:pinning}. We note, that the ansatz~\eqref{eq:circular-dw} does not takes into account any specific characteristics of the systems and might underestimate the pinning strength. The total energy shown in~Fig.~\ref{fig:pinning}(a) is a sum of four terms, shown in~Figs.~\ref{fig:pinning}(c)--(f). Exchange and anisotropy energies (related to the coefficient $\mathscr{K}$) are monotonically increasing functions and cannot pin the skyrmion. The energy $\mathcal{E}_\text{\textsc{dmi}\,2}$, related to $\mathscr{D}_2$, shows a small maximum, which does not contribute significantly to the pinning effect. In contrast, the energy $\mathcal{E}_\text{\textsc{dmi}\,1}$, related to $\mathscr{D}_1 \propto \varkappa_1$, exhibits a pronounced minimum near $R_1$, resulting in the appearance of a local minimum in the total energy. 

We estimate the energy gap of $\Delta \mathcal{E} = 750$~K for the case of a nanoindentation of Co/Pt stacks ($A = 10 $~pJ/m, $K_u = 0.3 $~MJ/m$^3$, $M_\textsc{s} = 480$ kA/m, thickness of the Co layer $\mathcal{h} = 0.6$~nm) with a geometry considered in Fig.~\ref{fig:pinning}.

\section{Discussion}
\label{sec:discussion}

The geometry of a ferromagnetic film makes a significant impact on static and dynamic skyrmion properties. Finite dimensions of nanostructures can lead to the confinement of a skyrmion~\cite{Rohart13} and skyrmion formation under an external influence~\cite{Jiang16}. Considering curvilinear effects, an alternative way to stabilize skyrmions is to utilize a curvature-induced DMI of interfacial type in samples with geometrically defined axis of anisotropy~\cite{{Gaididei14}, {Sheka15},Kravchuk16a,Kravchuk18a}.  

In this work we studied magnetic nanoindentations of radial symmetry with locally perpendicular easy axis of magnetization. We propose a coordinate transformation allowing to incorporate metric of the curved surface into spatially dependent material parameters and obtain a driven skyrmion equation with the left-hand side in the form of a well-known description of flat systems, see e.g.~\cite{Leonov16}. Our approach uncovers two mechanisms of skyrmion stabilization. 

(i) The first mechanism is based on the appearance of the geometry-induced Dzyaloshinskii--Moriya interaction~\cite{Kravchuk16a}. The consequence of this effect is the possibility to form \emph{small sized} skyrmions in the region of maximal curvature. 

(ii) The second mechanism, addressed in this work, governs the skyrmion size by the curvature gradient, which results in \emph{tunable-size} skyrmions. 

In both these cases skyrmions are static solutions because their structure is determined by the distribution of material parameters. The size of skyrmions of the type (i) is limited by the characteristic magnetic length $\ell$ due to their localization in the region of the curvature maximum, where the curvature is approximately constant~\cite{Kravchuk16a}. The spatial inhomogeneity of the DMI and anisotropy coefficients becomes crucial when considering the magnetization textures of type (ii).

In the case of one-dimensional systems the local change in the anisotropy is the source of a domain wall nucleation~\cite{Kronmueller87a} and attractive or repulsive pinning in a magnet~\cite{Skomski03}. The stabilization of circular domain walls usually appears due to magnetostatics while the size can be also governed by the inhomogeniety of the anisotropy ~\cite{Sapozhnikov15a,Sapozhnikov16, Zhang16b}. In a curvilinear nanomembrane two effective anisotropies and two effective DMI appear and each of them is related to the corresponding principal curvature~\cite{Gaididei14,Sheka15}. For the radially-symmetric textures considered in this work, one of the curvature-induced anisotropies disappears due to the symmetry of the object and texture, while the sum of the intrinsic anisotropy and the second geometry-induced one is incorporated in the coefficient $\mathscr{K}$ in Eq.~\eqref{eq:static-equation}, see expression~\eqref{eq:K-eff}. 

The curvature-driven skyrmion of a large radius appears in the following way. If a circular domain wall is nucleated, e.g. due to magnetic field along $\hat{\vec{z}}$ axis, it is located on a bend of the magnetic nanomembrane as a result of the localized DMI. While the coordinate-dependent anisotropy only changes the slope of the energy landscape [Fig.~\ref{fig:pinning}(d)], the first DMI coefficient $\mathscr{D}_1$ depends only on the bend parameters, see \eqref{eq:D1-D2}, and reduces the N\'{e}el domain wall energy of the preferred chirality during the domain wall positioning on the bend. The second DMI coefficient $\mathscr{D}_2$, see \eqref{eq:D1-D2}, also can lower the domain wall energy, but it is inversely proportional to the radius of the nanoindentation and does not significantly affect the magnetization texture  (in the case of large objects). Note, that the crucial role is played by a spatially localized distribution of $\mathscr{D}_1$ only, whose gradient results in a local energy minimum for a skyrmion with respect to its radius, see Fig.~\ref{fig:pinning}. 

The skyrmion radius is determined by the relation of the DMI constant to the domain wall energy density in flat systems. Therefore, large radius skyrmions can be described by a circular domain wall ansatz~\cite{Rohart13, Kravchuk18}. In contrast to this case, a bend of a ferromagnetic nanomembrane provides a pinning potential for a circular domain wall separating flat and inclined parts. Then, the size of a chiral texture is determined both by the area enclosed by a bend and effective DMI exceeding the given critical value. This is similar to the appearance of one-dimensional chiral domain walls whose energy decreases proportionally to the DMI constant~\cite{Heide08, Rohart13}. 

Using a model of a circular curved nanomembrane projected to a plane~\eqref{eq:rho-via-r}, we demonstrate a good agreement with the exact numerical calculations for the bend of a finite width, see Fig.~\ref{fig:stability} and Appendix~\ref{app:sk-stab} for details. 
Our analytical model takes into account only the first curvature-induced DMI $\mathscr{D}_1$ in the form of a Dirac $\delta$-function neglecting the second DMI coefficient, $\mathscr{D}_2$, and with the constant anisotropy $\mathscr{K}_0 = 1$. The coordinate-dependent coefficient $\mathscr{D}_1$ determines the skyrmion radius $R_\text{sk}$. Note, that in this case, the skyrmion equation~\eqref{eq:static-equation} differs from the planar case only by the presence of a driving force~\eqref{eq:force} appearing after the energy variation. Therefore, this allows us to predict the same properties of circular N\'{e}el domain walls in planar films with nonzero DMI in a narrow circular region: N\'{e}el domain walls should be pinned in the region with the DMI sign selecting the magnetization rotation direction inward or outward.

The estimation of the pinning strength for a skyrmion formed in a Co/Pt-based nanoindentation with $\varkappa_0 = 0.25$ and $w = 1.5$ (bending angle $\alpha \approx 26^\circ$ with the bend width of about 15~nm) shown in Fig.~\ref{fig:pinning} leads to the energy gap of about $ 10^3$~K stabilizing skyrmion on a bend. This model refers to actively studied nanopatterned media including circular nanoindentations~\cite{Makarov07,Makarov09a,Schulze10,Neu13} and convex structures like cones \cite{Ball14,Ball17}, caps~\cite{Ulbrich06,Streubel12,Streubel12a} and spheres~\cite{Kravchuk16a}. One can compare this prediction with the experiments for caps and nanoindentations~\cite{Makarov07,Brombacher09,Schulze10,Neu13} covered with perpendicularly magnetized Co/Pt multilayers. 
The presence of experimentally observed single domain features localized in curved regions is typically attributed to a (partial) exchange decoupling between the magnetic nanostructure and a flat film. Here, we show how the tilt of the anisotropy axis on a bend of a nanomembrane can give a significant contribution to the domain wall pinning due to the geometry-induced effective DMI, which is strongly localized in the bend area, see Fig~\ref{fig:pinning}. We speculate that these objects can be large size skyrmions discussed in this work.

\section{Acknowledgements}

O.~V.~P. and D.~D.~S. thank Helmholtz-Zentrum Dresden-Rossendorf e.~V. (HZDR), where part of this work was performed, for their kind hospitality and acknowledge the support from the Alexander von Humboldt Foundation (Research Group Linkage Programme). V.~P.~K. acknowledges the support from the Alexander von Humboldt Foundation. O.~V.~P. acknowledges the support from DAAD (code No. 91530902). This work was financed in part via the BMBF project GUC-LSE (federal research funding of Germany FKZ: 01DK17007), German Research Foundation (DFG) Grant MA 5144/9-1, and National Academy of Sciences of Ukraine, Project No. 0116U003192. A.~S. was supported by the U.S. Department of Energy. 

\appendix


\section{Surfaces of Revolution}
\label{app:surf-rev}

We form the surface $\vec{\varsigma}$ by the revolution of a curve $\vec{\gamma} = r\hat{\vec{x}} + z\hat{\vec{z}}$ around $\hat{\vec{z}}$: $\vec{\varsigma}(r,\chi) = \{r\cos\chi, r\sin\chi, z(r) \}$. Here and below, all distances are measured in the units of the magnetic length $\ell$. The complete definition of geometrical properties of $\vec{\varsigma}$ can be inferred trough two principal curvatures,
\begin{equation}
\varkappa_1(r) = \dfrac{\partial^2_{rr}z}{Z^3},\quad \varkappa_2(r) = \dfrac{\partial_rz}{rZ}
\end{equation}
with $Z(r) = \sqrt{1+(\partial_r z)^2}$.
Note, that the first principal curvature coincides with the curvature of the generatrix $\vec{\gamma}$. In the case of a surface of revolution, there is a relation $\varkappa_1 = \partial_r (r \varkappa_2)$. The way to extend $\vec{\varsigma}$ along the normal $\vec{n}$ without self intersection in the surface vicinity is to introduce coordinates along the principal directions (radial and polar directions), $\vec{e}_1 = \partial_r \vec{\varsigma}/| \partial_r \vec{\gamma}|$ and $\vec{e}_2 = \{ -\sin\chi, \cos\chi, 0 \}$. Then the normal is $\vec{n} = \vec{e}_1 \times \vec{e}_2$. The area element is $\mathrm{d}S = rZ\mathrm{d}r\mathrm{d}\chi$.


\section{Energy of a Curvilinear Ferromagnetic Nanomembrane}
\label{app:curv-energy}

The energy of a ferromagnetic nanomembrane reads
\begin{equation}\label{eq:energy-total}
E = h \bigintssss \left[ A \mathscr{E}_\text{ex} - K_u m_n^2 - \dfrac{M_\textsc{s}}{2} (\vec{m}\cdot \vec{H}_\text{d}) \right] \mathrm{d}S,
\end{equation}
where $A$ is the exchange stiffness, $K_u>0$ is the constant of uniaxial anisotropy, $m_n$ is the normal magnetization component and $\vec{H}_\text{d}$ is the demagnetizing field. In a thin film limit $h \lesssim \ell$ we incorporate the magnetostatic effects in the effective anisotropy $K = K_u - 2\pi M_\textsc{s}^2$~\cite{Gioia97,Carbou01,Kohn05a,Fratta16b}. 

The exchange energy density for an angular parametrization of the magnetization texture $\vec{m} = \sin\theta \cos\phi\, \vec{e}_1 + \sin\theta \sin\phi\, \vec{e}_2 + \cos\theta\, \vec{n}$ reads \cite{Gaididei14,Sheka15}: 
\begin{equation} \label{eq:exchange-shell-angular}
\mathscr{E}_{\text{ex}} = 
\left[\vec{\nabla}\theta  - \vec{\varGamma }\right]^2\!\! + \left[\sin\theta \left(\vec{\nabla}\!\phi-\vec{\varOmega }\right)\! - \!\cos \theta \partial_\phi \vec{\varGamma }\right]^2
\end{equation}
with $\vec{\varOmega}$ being a spin connection with components $\varOmega_\mu = \vec{e}_1\cdot \nabla_\mu  \vec{e}_2$ for $\mu = 1,2$ and the vector  $\vec{\varGamma}(\phi)= \varkappa_1 \cos\phi\, \vec{e}_1 + \varkappa_2 \sin\phi\, \vec{e}_2 $. Here, $\vec{\nabla} = \frac{\vec{e}_1}{Z} \partial_r + \frac{\vec{e}_2}{r}\partial_\chi$ 
denotes a surface del operator in its curvilinear form. In particular, for surfaces of revolution $\varOmega_1 = 0$ and $\varOmega  := \varOmega_2  = -1/(rZ)$.

The static Landau--Lifshitz equation have the following form
\begin{subequations} \label{eq:LLE}
\begin{align}  \label{eq:LLE-1}
\begin{aligned}
& \vec{\nabla}^2 \theta - \sin\theta \cos\theta \left[1 - \left(\partial_\phi \vec{\varGamma }\right)^2 + \left(\vec{\nabla}\phi - \vec{\varOmega}\right)^2\right]\\
& +\cos 2\theta \left(\vec{\nabla}\phi - \vec{\varOmega}\right)\cdot\partial_\phi \vec{\varGamma }  - \vec{\nabla} \cdot \vec{\varGamma}=0,
\end{aligned}
\\
\label{eq:LLE-2} %
\begin{aligned}
& \vec{\nabla} \cdot \left(\sin^2\theta \vec{\nabla}\phi\right) + \sin^2\theta \left[ \left(2\vec{\nabla}\theta - \vec{\varGamma}\right)\cdot \partial_\phi \vec{\varGamma} - \vec{\nabla} \cdot \vec{\varOmega}\right]\\
& - \sin\theta \cos\theta \left[2\vec{\nabla}\theta \cdot \vec{\varOmega} + \vec{\nabla} \cdot \partial_\phi \vec{\varGamma} + \left(\vec{\nabla}\phi - \vec{\varOmega}\right) \cdot \vec{\varGamma}\right] = 0.
\end{aligned}
\end{align}
\end{subequations}
There exists a class of azimuthally symmetric solutions
\begin{equation} \label{eq:theta(s)}
\theta=\theta(r), \qquad \phi=0,\pi, \; \cos\phi = \pm1.
\end{equation} 
For this class of solutions, it is convenient to use another angular parametrization:
\begin{equation} \label{eq:Theta(s)}
\vec{m} = \sin\Theta(r) \vec{e}_1 + \cos\Theta(r) \vec{n}, \qquad \Theta\in\mathbb{R}.
\end{equation}
The energy density~\eqref{eq:exchange-shell-angular} reads
\begin{equation}\label{eq:exch-theta}
\begin{aligned}
\mathscr{E}_\text{ex} & = \frac{(\partial_r\Theta)^2}{Z^2} - \frac{2\varkappa_1\partial_r\Theta }{Z} + \varkappa_1^2 + \varkappa_2^2 \\
& + (\varOmega^2 - \varkappa_2^2)\sin^2\Theta + 2 \varOmega \varkappa_2 \sin\Theta\cos\Theta.
\end{aligned}
\end{equation}

\begin{figure*}
\includegraphics[width=\linewidth]{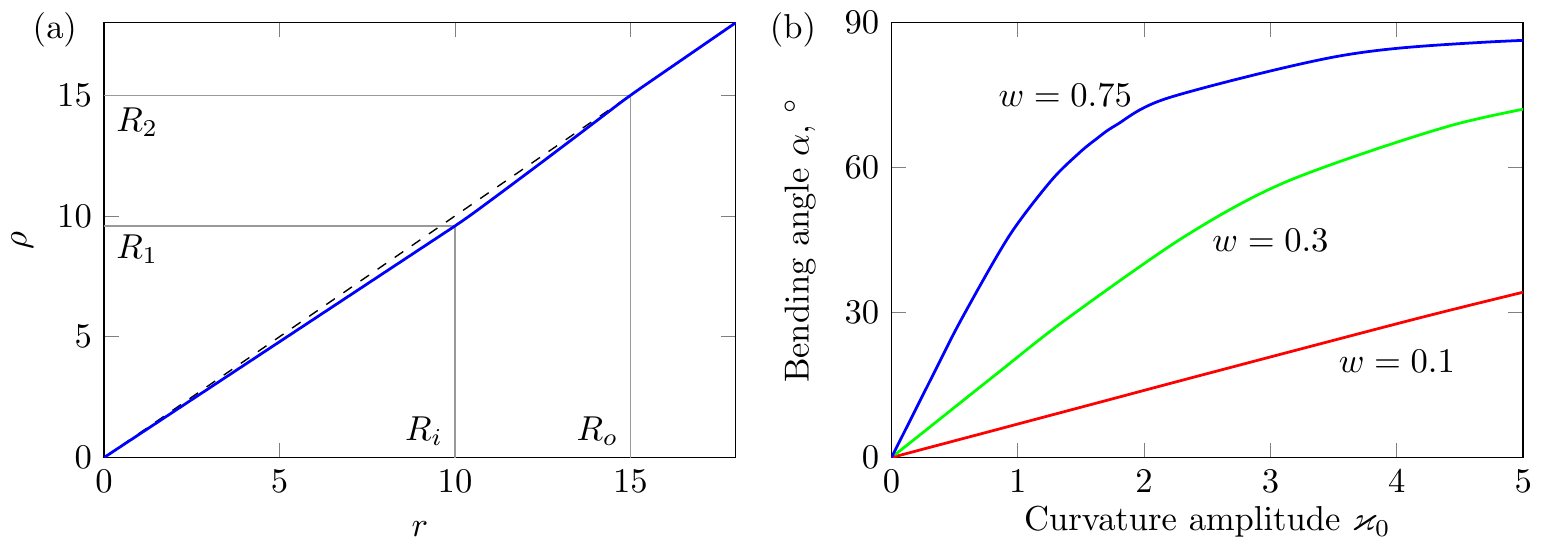}
\caption{\textbf{Characteristics of the nanomembrane geometry.} (a) A relation between the distance to the axis $r$ of the sample and the surface polar coordinate $\rho$ according to~\eqref{eq:rho-via-r}. The dependence $\rho(r)$ is shown with a line, bisecting the quadrant, is shown with a dashed line. The Plot is built for parameters taken in Fig.~\ref{fig:intro}. Here, $R_1 \approx 9.59$ and $R_2 \approx 14.99$. (b) Bending angle $\alpha$ for different spatial localizations $w$ of the curvature.}
\label{fig:rho-plot}
\end{figure*}

We start with the case of an isotropic magnet, $K = 0$. The ground state of the model is the homogeneous (in the physical space) state. We limit ourselves by the homogeneous magnetization distributions along $z$-axis with $\vec{m}_\text{h} = \mathcal{C}_h \hat{\vec{z}}$:
\begin{equation} \label{eq:hom-state}
\begin{split}
\cos\Theta_\text{h} &= \frac{\mathcal{C}_h}{Z}, \quad  \sin\Theta_\text{h} = \frac{\mathcal{C}_h\partial_rz}{Z}
\end{split}
\end{equation}
with $\mathcal{C}_h = \pm 1$.

It is instructive to represent the energy functional  \eqref{eq:exch-theta} in terms of the angle $\psi$, which characterizes the deviation  from the homogeneous state
\begin{equation} \label{eq:deviation}
\Theta(r) = \Theta_\text{h}(r) + \psi(r).
\end{equation}
Then, the energy density reads
\begin{equation} \label{eq:energy-skyrmion-ex}
\mathscr{E}_\text{ex} = \frac{(\partial_r\psi)^2}{Z^2} + \frac{\sin^2\psi}{r^2}
\end{equation}
The static equation can be written as
\begin{equation} \label{eq:static}
\frac{r}{Z}\frac{\mathrm{d}}{\mathrm{d}r} \left(\frac{r}{Z}\frac{\mathrm{d}\psi}{\mathrm{d}r}\right) - \sin\psi\cos\psi = 0
\end{equation}
The change of the variable~\eqref{eq:rho-via-r} with the limiting values $\rho(0) = 0$ and $\rho(\infty) = \infty$ allows to rewrite~\eqref{eq:static} as
\begin{equation} \label{eq:static-rho}
\psi'' + \frac{1}{\rho} \psi' - \frac{1}{\rho^2}\sin\psi\cos\psi=0,
\end{equation}
This expression coincides with the equation of radially symmetric magnetic texture for an isotropic planar ferromagnet~\cite{Kosevich90}. 
Fig.~\ref{fig:rho-plot}(a) shows the relation between the surface polar coordinate $\rho$ and the distance to the symmetry axis $r$.

We look for a skyrmion solution of the equation \eqref{eq:static}, which satisfies the boundary conditions
\begin{equation} \label{eq:bcs}
\psi(r=0)=\psi(\rho=0) = \pi, \quad \psi(r=\infty) = \psi(\rho=\infty) = 0.
\end{equation}
The corresponding solution of Eq.~\eqref{eq:static-rho} is the well--known Belavin--Polyakov skyrmion solution \cite{Belavin75}
\begin{equation} \label{eq:BP}
\begin{split}
\tan\frac{\psi_{\textsc{bp}}}{2} &= \frac{R}{\rho}, \quad R=\text{const},\\
\Theta_{\textsc{bp}}(r) &= \Theta_\text{h}(r) + \psi_{\textsc{bp}}(r).
\end{split}
\end{equation}
The energy of the Belavin--Polyakov skyrmion $E = 8\pi A t$ does not depend on its radius, which results in the skyrmion instability. An efficient way of static stabilization of the skyrmion structure is to take into account both anisotropy and DMI. In our case both interactions appear effectively due to the curvature of the nanomembrane.

\begin{figure}
\includegraphics[width=\linewidth]{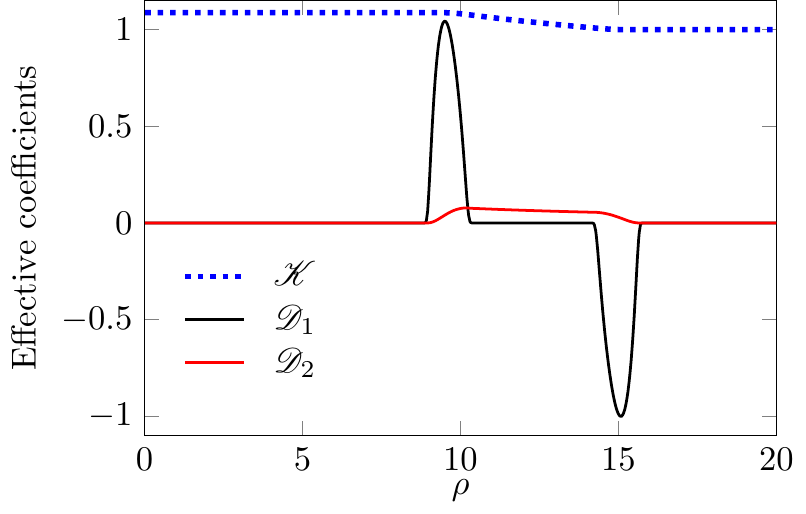}
\caption{\textbf{Effective coefficients of the energy of the radially symmetric solution.} Effective anisotropy~\eqref{eq:K-eff} and two DMI coefficients~\eqref{eq:D1-D2} as a function of the $\rho$ coordinate for the sample parameters shown in Fig.~\ref{fig:intro}.}
\label{fig:matparams}
\end{figure}

The energy~\eqref{eq:energy-total} for the radially symmetric solution~\eqref{eq:Theta(s)} up to a constant reads
\begin{equation}\label{eq:energy-renorm}
\mathcal E = \dfrac{E}{2\pi At} = \int \mathscr{E} \rho\mathrm d\rho
\end{equation}
with the energy density
\begin{equation} \label{eq:energy}
\begin{aligned}
\mathscr{E}=&
\underset{\text{exchange}}{\underbrace{\Theta'^2 + \frac{\sin^2\Theta}{\rho^2}}} + \underset{\text{anisotropy}}{\underbrace{\mathscr{K} \sin^2\Theta\vphantom{\frac{\sin^2\Theta}{\rho^2}}}}\\
&
\underset{\text{DMI\,1}}{\underbrace{-\mathscr{D}_1\Theta'\vphantom{\dfrac{\sin\Theta}{\rho}}}} \quad \underset{\text{DMI\,2}}{\underbrace{- \mathscr{D}_2 \frac{\sin\Theta \cos\Theta}{\rho}}},
\end{aligned}
\end{equation}
where coefficients are functions of the principal curvatures of the nanomembrane $\varkappa_1(\rho)$ and $\varkappa_2(\rho)$. The parameter $\mathscr{K}$ can be interpreted as an effective anisotropy parameter
\begin{subequations}\label{eq:effective-parameters}
	\begin{equation} \label{eq:K-eff}
	\mathscr{K} = \frac{r^2}{\rho^2} \left(1 -2\varkappa_2^2 \right)
	\end{equation}
	with $r = r(\rho)$ here and below. Parameters $\mathscr{D}_1$ and  $\mathscr{D}_2$  can be treated as parameters of the effective DMI interaction:
	\begin{equation} \label{eq:D1-D2}
	\mathscr{D}_1 = \frac{2 r \varkappa_1}{\rho}, \qquad \mathscr{D}_2 = 2r' \varkappa_2,
	\end{equation}
\end{subequations}
An example of the spatial distribution of $\mathscr{K}\!$, $\mathscr{D}_1$ and $\mathscr{D}_2$ is shown in Fig.~\ref{fig:matparams}.

There is a striking correspondence between \eqref{eq:energy} and the energy density of a chiral skyrmion in a planar magnet with an intrinsic DMI~\cite{Leonov16}. The only difference is the coordinate dependence of the anisotropy $\mathscr{K}$ and two DMI constants $\mathscr{D}_{1,2}$. An example of the dependence of these coefficients on $\rho$ is shown in Fig.~\ref{fig:matparams}. The first DMI coefficient $\mathscr{D}_1\propto\varkappa_1(\rho)$. The second DMI coefficient $\mathscr{D}_2\propto \varkappa_2$ is nonzero only at the tilted part of the nanoindentation.


\section{Model of a Nanoindentation}
\label{app:sk-stab}

We model a nanoindentation with a flat inner part considering a sharp indent of a conic frustum shape
\begin{equation}\label{eq:frustum}
z_\text{sh}(r) = 
\begin{dcases*}
0, & when $r<R_i$\\
(r-R_i)\tan\alpha,& when $R_i\le r \le R_o$\\
z_0, & when $r>R_o$
\end{dcases*}
\end{equation}
with $\tan \alpha = z_0/(R_o-R_i)$. Using the mapping condition~\eqref{eq:rho-via-r}, we obtain
\begin{equation}\label{eq:fr-r-by-rho}
r(\rho) = 
\begin{dcases*}
\frac{R_i}{R_1}\rho, & when $\rho<R_1$,\\
R_2 \left(\frac{\rho}{R_2}\right)^{\cos\alpha},& when $R_1<\rho<R_2$,\\
\rho, & when $\rho>R_2$
\end{dcases*}
\end{equation}
with $R_1 = R_o(R_i/R_o)^{\sec \alpha} < R_2$ and $R_2 = R_o$. Curvatures of the nanomembrane are given by~\eqref{eq:delta-curvature} and
\begin{equation}\label{eq:second-curv-rho}
\varkappa_2(\rho) = 
\begin{dcases*}
0,& when $\rho < R_1$ and $\rho > R_2$,\\
\dfrac{\alpha \cos\alpha}{R_2} \left(\dfrac{R_2}{\rho}\right)^{\cos\alpha}, & when $R_1 < \rho<R_2$,\\
\dfrac{\alpha \cos(\alpha/2)}{2R_1}, & when $\rho = R_1$,\\
\dfrac{\alpha\cos(\alpha/2)}{2R_2}, & when $\rho = R_2$.
\end{dcases*}
\end{equation}

We assume, that both $R_1$ and $R_2$ are large enough to omit all terms of the order $1/\rho$ and higher in the energy density~\eqref{eq:energy}. Then, the expression~\eqref{eq:energy} can be written as
\begin{equation} \label{eq:energy-dens-toy}
\begin{aligned}
\mathscr{E}_\text{0}=& \Theta'^2 + \mathscr{K}_0 \sin^2\Theta\\
&
- 2\alpha\dfrac{r}{\rho} \left[ \dfrac{R_1}{R_i}\delta(\rho-R_1) - \delta(\rho - R_2) \right]\Theta',
\end{aligned}
\end{equation}
where 
\begin{equation}\label{eq:}
\mathscr{K}_0 = 
\begin{dcases*}
\left(\dfrac{R_2}{R_i}\right)^{2(\sec\alpha-1)},& when $\rho < R_1$,\\
\left(\dfrac{R_2}{\rho}\right)^{2(1-\cos\alpha)}, & when $R_1 \le \rho \le R_2$,\\
1, & when $\rho > R_2$.
\end{dcases*}
\end{equation}
This expression can be simplified to $\mathscr{K}_0 \approx 1$ in a wide range of ratios $R_1/R_2$ for small angles $\alpha$ and in a wide range of $\alpha$ if inner and outer radii $R_1$ and $R_2$ are comparable, see Fig.~\ref{fig:matparams}.

Applying here the circular domain wall ansatz~\eqref{eq:circular-dw}, we obtain the total energy~\eqref{eq:skyrmion-en}. The condition for the existence of a local minimum in~\eqref{eq:skyrmion-en} gives expressions~\eqref{eq:dw-stab-stabo}.

In a general case of a curvature with a finite localization, we can smoothen the  shape~\eqref{eq:frustum} as
\begin{equation}\label{eq:frustum-smooth}
z(r) = \int_{-w}^{\min\{r,w\}} g_w(\zeta) z_\text{sh}(r-\zeta)\mathrm{d}\zeta,
\end{equation}
where the mollifier $g_w(r)$ reads
\begin{equation}\label{eq:mollifier}
g_w(\zeta) = 
\begin{dcases*}
\dfrac{C_g}{w} \exp \left( \dfrac{w^2}{\zeta^2-w^2} \right) & when $-w \le \zeta \le w$,\\
0 & otherwise.
\end{dcases*}
\end{equation}
Here, $C_g  \approx 2.25$ from the condition $\int_{-w}^{w}g_w(\zeta)\mathrm{d}\zeta = 1$. Mollifying~\eqref{eq:frustum-smooth} guarantees that the first principal curavture $\varkappa_1(r)$ is nonzero only in a region of $2w$ around the bend. Bending angle as a function of $w$ and curvature amplitude $\varkappa_0$ is shown in Fig.~\ref{fig:rho-plot}(b). All lines asymptotically tend to $\alpha = 90^\circ$. A stability diagram shown in Fig.~\ref{fig:stability} is built using the stability analysis described in Section~IV of~\cite{Kravchuk18}.

%

%
\end{document}